\documentclass[twocolumn, showpacs]{revtex4}

\usepackage{graphicx}
\begin{document}
\bibliographystyle{apsrev}

\title{First observation of the acceleration of a single bunch by using the induction device in the KEK Proton Synchrotron}

\author{Ken Takayama $^{1, 2, 3}$}
\author{Kunio Koseki $^{2}$}
\author{Kota Torikai $^{1, 4}$}
\author{Akira Tokuchi $^{5}$}
\author{Eiji Nakamura $^{1, 2}$}
\author{Yoshio Arakida $^{1}$}
\author{Yoshito Shimosaki $^{1}$}
\author{Masayoshi Wake $^{1}$}
\author{Tadaaki Kouno $^{1}$}
\author{Kazuhiko Horioka $^{3}$}
\author{Susumu Igarashi $^{1}$}
\author{Taiki Iwashita $^{1}$}
\author{Atsushi Kawasaki $^{5}$}
\author{Jun-ichi Kishiro $^{1, 6}$}
\author{Makoto Sakuda $^{7}$}
\author{Hikaru Sato $^{1}$}
\author{Makoto Shiho $^{3, 6}$}
\author{Masashi Shirakata $^{1}$}
\author{Tsuyoshi Sueno $^{1}$}
\author{Takeshi Toyama $^{1}$}
\author{Masao Watanabe $^{6}$}
\author{Isao Yamane $^{1}$}

\affiliation{$^1$ Accelerator Laboratory, High Energy Accelerator Research Organization (KEK), 1-1 Oho, Tsukuba, Ibaraki 305-0801, Japan}
\affiliation{$^2$ The Graduate University for Advanced Studies, Hayama, Miura, Kanagawa 240-0193, Japan}
\affiliation{$^3$ Tokyo Institute of Technology, Nagatsuta 4259, Yokohama 226-8502, Japan}
\affiliation{$^4$ Department of Applied Quantum Physics and Nuclear Engineering, Faculty of Engineering, Kyushu University, 6-10-1 Hakozaki, Fukuoka-shi, Fukuoka 812-8581, Japan}
\affiliation{$^5$ Nichicon (Kusatsu) Corporation, 2-3-1 Yagura, Kusatsu, Shiga 525-0053, Japan}
\affiliation{$^6$ Japan Atomic Energy Research Institute, 2-4 Shirane, Tokai, Naka, Ibaraki 319-1195, Japan}
\affiliation{$^7$ Physics Department, Okayama University, 3-1-1 Tsushimanaka, Okayama, Okayama 700-8530, Japan}

\date{\today}

\begin{abstract}
A single RF bunch in the KEK proton synchrotron was accelerated with an induction acceleration method from the injection energy of 500 MeV to 5 GeV.
\end{abstract}

\maketitle



\begin{figure}
\includegraphics{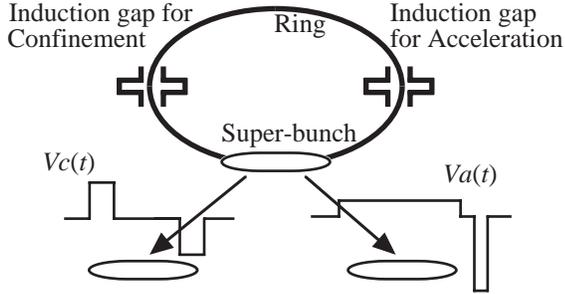}
\caption{Schematic view of the induction synchrotron and step-voltage profiles for confining and acceleration.}
\label{fig1}
\end{figure}

Four years ago, the concept of an induction synchrotron employing induction accelerating devices was proposed by Takayama and Kishiro \cite{Ref1} for the purpose of overcoming the shortcomings, such as a limitation of the longitudinal phase-space available for the acceleration of charged particles in an RF synchrotron, which has been one of indispensable instruments for nuclear physics and high-energy physics since the invention by McMillan \cite{Ref2} and Veksler \cite{Ref3}.  Accelerating devices in a conventional synchrotron, such as an RF cavity, are replaced with induction devices in the induction synchrotron.  A gradient focusing force, seen in the RF waves, is not indispensable for the longitudinal confinement of particles.  Pulse voltages, which are generated at both edges of some time-period with opposite sign, as shown in  Fig.~\ref{fig1}, are capable of providing longitudinal focusing forces.  A pair of barrier-voltage pulses work in a similar way to the RF barrier, which has been demonstrated at FNAL and BNL \cite{Ref4}.  The acceleration and longitudinal confinement of charged particles are independently achieved with induction step-voltages in the induction synchrotron.  This notable property of the separated-function in the longitudinal direction brings about a significant freedom of beam handling never seen in a conventional RF synchrotron, in which radio-frequency waves in a resonant cavity simultaneously take both roles of acceleration and longitudinal confinement.  Associated with this separated-function, various figure of merits are expected.  The formation of a super-bunch, which is an extremely long bunch with a uniform line-density, and the use of which is considered in a proton driver for the second-generation neutrino physics and a future hadron collider \cite{Ref5}, is most attractive.  In addition, transition crossing without any longitudinal focusing seems to be feasible \cite{Ref1}, which could substantially mitigate undesired phenomena, such as bunch shortening due to non-adiabatic motion and microwave instabilities \cite{Ref6}.

The time-duration between barrier-voltage pulses determines the size of a super-bunch.  For accelerating a super-bunch, a long accelerating step voltage with a pulse length on the order of $\mu$sec is required.  The voltage has to be generated at the revolution frequency of the beam in a ring.  In the case of a ring-circumference of 300 m, a repetition rate of 1 MHz is required.  So far, there has been no induction accelerating system capable of meeting these parameters.  Recently, prototype devices, which can generate a 250nsec flat-top voltage at a repetition rate of cw 1 MHz, have been assembled at KEK after the 3-year R\&D stage, and combined with the existing RF accelerating system.  A single proton bunch trapped in an RF bucket was accelerated with the induction accelerating system from 500 MeV to 5 GeV during 1.6 sec.  In this letter, we report on the first experimental result of the induction acceleration of a single RF bunch in the KEK proton synchrotron (KEK-PS) as well as a brief description of the induction accelerating system.



\begin{figure}
\includegraphics{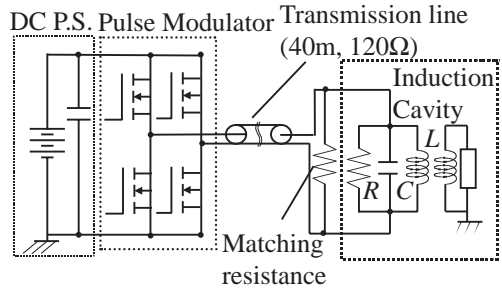}
\caption{Equivalent circuit for the induction accelerating system.}
\label{fig2}
\end{figure}

The key devices required to realize an induction synchrotron are an induction accelerating cavity \cite{Ref7} and a pulse modulator driving the cavity \cite{Ref8}.  These devices are notably different from similar devices employed in modern linear induction accelerators \cite{Ref9}.  Remarkable characteristics are its switching characteristics, repetition ratio, and duty factor.  Another different feature is that the pulse modulator has to be kept far from the induction cavity placed in the accelerator tunnel, because the solid-state power switching elements obtainable at present can't survive an extremely high radiation dose.  Thus, the pulse modulator is connected with the accelerating cavity through a long transmission cable.  In order to reduce reflection from the load, a matching resistance has been installed at the end of the transmission cable.  The induction accelerating system consists of an induction cavity with a matching load, a transmission cable, a pulse modulator, and a DC power supply.  A typical system capable of generating a step-pulse of 2 kV output voltage and 18 A output peak current at 1 MHz with 50\% duty has been demonstrated at KEK.  An equivalent circuit for the induction accelerating system is shown in Fig.~\ref{fig2}.

The core material of the induction cavity employed for the first acceleration experiment is a nanocrystalline alloy, called Finemet (Hitachi Metal).   Heat generated due to core loss is cooled down by insulation oil.  The electrical parameters of each unit-cell are given by the capacitance of 260 pF, the inductance of 110 $\mu$H and the resistance of the 330 $\Omega$, which determine the properties of pulse rising and falling.  Three unit cells with a 2 kV output voltage per unit are mechanically combined into a single module for the convenience of installation.  Since the inner conductor with three ceramic gaps is common to three unit-cells, but both sides of each gap are electrically connected to the outer edges of each cell, a particle is accelerated with the same voltage passing these gaps.

A full-bridge switching circuit in the pulse-modulator, which is depicted in Fig.~\ref{fig2}, was employed because of its simplicity.  The pulse modulator is capable of generating bipolar rectangular shaped voltage pulses.  The full-bridge type pulse-modulator consists of four identical switching arms.  Each switching arm is composed of 7 MOS-FETs, which are arranged in series.  Their gates are driven by their own gate-driving circuits.  The gate signals are generated by converting light signals provided from the pulse controller, which is a part of the accelerator control system, to electronic signals.  Its details will be presented elsewhere \cite{Ref8}.

\begin{figure}
\includegraphics{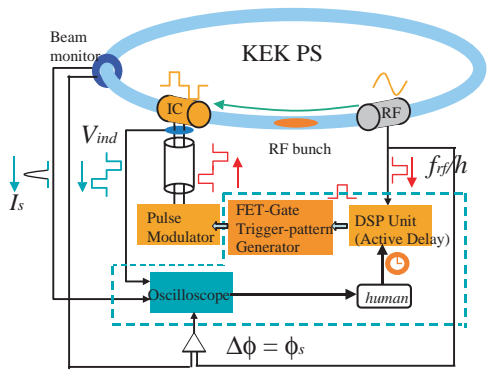}
\caption{Schematic view of the hybrid accelerating system.}
\label{fig3}
\end{figure}

The entire system employed for the current experiment is schematically shown in Fig.~\ref{fig3}.  The generation of a 2 kV voltage pulse is directly controlled by trigger pulses for the switching elements of the pulse modulator, the master signal of which is created in the DSP synchronizing with ramping of bending magnets, and gate-driving signal patterns initiated by this master trigger-signal are generated by the following signal-pattern generator to be sent to the gate controller of the pulse modulators through a long coaxial cable.  The DSP counts the B-clock signal, and achieves the desired revolution frequency.  Any delay between the accelerating pulse and a bunch monitor signal is always corrected by the DSP.  The system is connected to the existing RF system through the RF signal, which shares the B-clock signal.  As a result, synchronized induction acceleration is guaranteed.  Here, the RF does not contribute to acceleration of the beam bunch, but gives the focusing force in the longitudinal direction; the beam bunch is in principle trapped around the phase of zero.  The machine parameters of the KEK PS employed for the experiment are as follows; transition energy $\gamma _t$ is 6.63, injection and extraction energy are 500 MeV/8 GeV, revolution frequency $f_0$ is 668 - 877 kHz, RF voltage $V_{rf}$ is 40 kV and harmonic number $h$ is 9.

\begin{figure}
\includegraphics{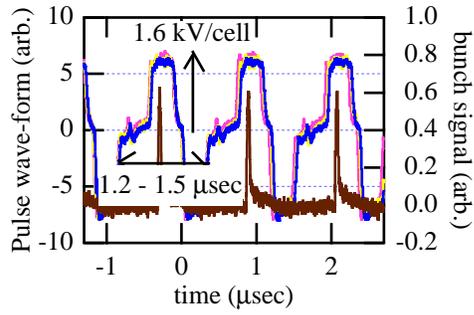}
\caption{Induction pulse wave-forms (pink, yellow, blue) and a bunch signal (brown).}
\label{fig4}
\end{figure}

In the experiment, the signals of the bunch monitor and three current transformers (CT), which always observed the current flow through the matching resistances, were monitored on a digital oscilloscope located at the central accelerator control room.  Before the experiment, an actual induced voltage at the ceramic gap, an output voltage of the transmission cable, and the CT signal were simultaneously measured and the correspondence between each other was well calibrated.  In addition, a delayed timing of the master gate signal triggering the pulse modulator is adjusted by the DSP so that the bunch signal would stay around the center of the induction voltage pulse through the entire accelerating period.  Typical wave-forms of the CT signals are shown in Fig.~\ref{fig4} together with the bunch signal.

\begin{figure}
\includegraphics{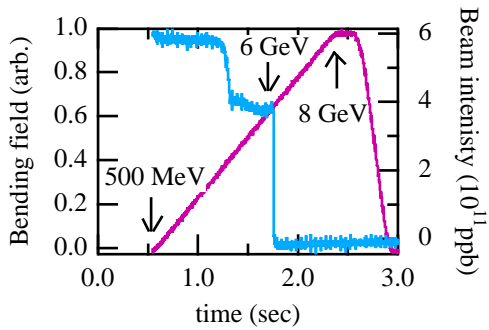}
\caption{Ramping pattern of the bending field (purple) and beam intensity (sky blue). The beam is stopped just below a transition at 1.75 sec, and some beam loss due to transverse motion is observed around 0.5 sec earlier before the transition.}
\label{fig5}
\end{figure}

Under coexistence with the RF voltage and the induction voltage, a charged particle receives an energy gain per turn,
\begin{equation}
eV_{acc}\left( t \right)=e\left[ {V_{rf}\sin \phi \left( t \right)+V_{ind}} \right],
\label{eq1}
\end{equation}
where $V_{rf}$ and $V_{ind}$ are the RF voltage and the induction voltage, respectively, and $\phi \left( t \right)$ is the position of the particle in the RF phase, $\omega _{rf} t$.  The orbit and energy of a particle are dominated by the following equations: for the force balance in the radial direction
\begin{equation}
m\gamma \cdot \frac{\left( {c\beta } \right)^2}{\rho}=ec\beta \cdot B\left( t \right),
\label{eq2}
\end{equation}
where $B \left( t \right)$ is the bending field and $\rho$ is the bending radius, and for the change in energy
\begin{equation}
mc^2\cdot \frac{d\gamma }{dt}=\frac{ec\beta }{C_0}\cdot V_{acc}\left( t \right).
\label{eq3}
\end{equation}
From Eq.~(\ref{eq2}) and (\ref{eq3}), the accelerating voltage must satisfy the relationship $V_{acc}\left( t \right)=\rho \cdot C_0\cdot dB/dt$,  so that the particle is synchronously accelerated with ramping of the bending field.  The bending field is linearly ramped over 1.7 sec, as shown in Fig.~\ref{fig5}.  In this linear ramping region, a $V_{acc}$ of 4.7 kV is required.  In the present experiment, the induction voltage was fixed to be 5.28 kV.

In order to confirm the induction acceleration, the phase signal, which shows the relative position of the bunch center to the RF, was measured through an entire accelerating region.  Particularly, we focused on three cases: (1) with an RF voltage alone, (2) with an RF voltage and a positive induction voltage, and (3) with an RF voltage and a negative induction voltage.  From Eq.~(\ref{eq1}), a theoretical prediction is $\phi _s=\sin^{-1} \left( {V_{acc}/V_{rf}} \right) \sim V_{acc}/V_{rf} =6.73$ degrees for case (1), $\phi _s=-0.68$ degree for case (2), $\phi _s= \sin^{-1} \left({2V_{acc}/V_{rf}} \right) \sim 2V_{acc}/V_{rf} =14.59$ degrees for case (3), where $\phi _s$ represents the position of the bunch center in the RF phase.  Since the induction voltage is devoted to the acceleration for case (2), the RF does not serve for acceleration, but takes a role of capturing alone; thus, the phase must be nearly zero.  In case (3), the RF has to give a two-times larger energy to the bunch than that in case (1) from the energy-conservation law; the phase should increase by a factor of two.  Actually, the time-evolution of the phases and a beam-current for each case has been observed.  There were notable differences between three cases for the temporal evolution in phase. The beam itself behaved in the same manner for the three cases as shown in Fig.~\ref{fig5}, where a temporal evolution of the beam intensity from 500 MeV to 5 GeV is shown.

\begin{figure}
\includegraphics{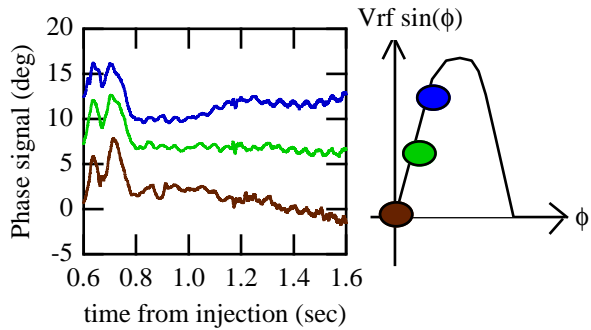}
\caption{Phase signals vs. time (sec) for case 1 (green), case 2 (brown), and case 3 (blue).}
\label{fig6}
\end{figure}

Before we proceed to an analysis of the experimental results, we have to pay attention to the phase signal amplitude, because there were several ambiguities in our phase pick-up monitoring system, such as time-varying offset and gain.  In order to avoid unknown factors, we have made a kind of calibration table, from which the relationship between the monitored phase pick-up voltage and the exact phase-angle is readable, by measuring the phase pick-up voltages corresponding to the different RF voltages from 40kV to 90kV.  In Fig.~\ref{fig6}, the experimental results are shown along with corrections obtained from a calibration table.  For case (1), the phase is very close to the theoretical prediction.  For case (2), the phase is nearly 0 as a rough estimate.  For case (3), the magnitude is higher by an about factor of two than that of Case (1).  At a first glance, we can find a qualitative agreement with the theoretical prediction.  Using the calibrated gain in the linear ramping region from 700 msec to 1.6 sec, we arrived at Table~\ref{table1}, where the results are more quantitatively compared with the prediction.

The experimentally obtained phases are slightly different from the theoretically predicted values.  A difference growing near the transition energy may be attributed to the fact that the RF bunch becomes shorter when approaching the transition energy; as a result, the effective acceleration voltage in this region sensitively depends on the relative position in the induction voltage with a droop.  Meanwhile, we realize that a difference between the accelerating voltage pulse-shape and the decelerating voltage pulse-shape, which can be seen in Fig.~\ref{fig4}, gives a slightly asymmetric evolution around Case (1).  From the over-all experimental results and their analysis, we conclude that a single RF bunch was accelerated from 500 MeV up to 5 GeV with the induction accelerating device, as expected.


The present induction voltage wave-form has a droop of 10-15\% over 250 nsec, although it has not been discussed.  To achieve a uniform acceleration of the super-bunch, flatness in the wave-form is important.  Several methods \cite{Ref11} to compensate the droop have been proposed.  They will be tried at the KEK PS.  Following the original scenario of the POP experiment to realize an induction synchrotron \cite{Ref1}, hereafter, we will proceed to its next step in the formation of a super-bunch by the induction step-barriers.


It is emphasized that for the first time charged particles in a high-energy accelerator ring were accelerated with an induction accelerating system.  The experimental fact of induction acceleration in a circular ring and the reality that the key devices developed for this purpose, such as a pulse modulator as a switching driver, are on our hands as big mile-stones for us to achieve an induction synchrotron in the near future, and a super-bunch hadron collider in the no-so-far future.  Last, it is emphasized that the induction system worked well through the entire operating period of 24 hours without any trouble.
 
The present research has been supported by the KEK PS division.  The authors acknowledge S. Ninomiya for providing much information concerning the KEK-PS RF system and helpful comments on the experimental results, and D. Arakawa for advice on the bunch monitor.  The present research has been financially supported by a Grant-In-Aid for Scientific Research for Creative Scientific Research (KAKENHI 15GS0217), and its early stage was partially supported by a Grant-In-Aid for Scientific Research on Priority Area (KAKENHI 140646221).

%
%
\begin{table}
\caption{Results quantitatively compared with the prediction.}
\label{table1}
\begin{tabular}{|l|c|c|}
\hline
  & experimental  (deg) & theoretical  (deg) \\ \hline
Case (1) & $6.5 \sim 7$ & 6.73\\ \hline
Case (2) & $2.0 \sim -2.0$ & -0.68\\ \hline
Case (3) & $10 \sim 13$ & 14.6\\ \hline
\end{tabular}
\end{table}
\end{document}